\documentclass[aps,prr,twocolumn,reprint,superscriptaddress]{revtex4-2}

\usepackage{hyperref}
\usepackage{graphicx}  
\usepackage{bm}        
\usepackage{amssymb}   
\usepackage{xspace}
\usepackage{verbatim}
\usepackage{color}
\usepackage{soul}
\usepackage{subfigure}
\usepackage[export]{adjustbox}
\usepackage{dcolumn}
\usepackage{amsthm,stmaryrd,mathtools} 
\usepackage{txfonts}
\usepackage{braket}
\usepackage{amsmath}
\usepackage{xcolor}
\usepackage{array}
\usepackage{multirow}
\usepackage{tabularx}
\usepackage{booktabs}
\usepackage{lipsum}

\begin{document}

\title{Voltage-controlled Hubbard spin transistor}

\author{Rozhin Yousefjani}%
\email{RozhinYousefjani@uestc.edu.cn}
\affiliation{Institute of Fundamental and Frontier Sciences, University of Electronic Science and Technology of China, Chengdu 610051, China}

\author{Sougato Bose}
\email{s.bose@ucl.ac.uk}
\affiliation{Department of Physics and Astronomy, University College London, Gower Street, London WC1E 6BT, United Kingdom}

\author{Abolfazl Bayat}
\email{abolfazl.bayat@uestc.edu.cn}
\affiliation{Institute of Fundamental and Frontier Sciences, University of Electronic Science and Technology of China, Chengdu 610051, China}

\begin{abstract}
Transistors are key elements for enabling computational hardware in both classical and quantum domains. Here, we propose a voltage-gated spin transistor using itinerant electrons in the Hubbard model which acts at the level of single electron spins. Going beyond classical spintronics, it enables the controlling of the flow of quantum information between distant spin qubits. The transistor has two modes of operation, open and closed, which are realized by two different charge configurations in the gate of the transistor. In the closed mode, the spin information between  source and drain is blocked while in the open mode we have free spin information exchange. The switching between the modes takes place within a fraction of the operation time which allows for several subsequent operations within the coherence time of the transistor. The system shows good resilience against several imperfections and opens up a practical application for quantum dot arrays.   	
\end{abstract}

\maketitle


\section{Introduction}
Transistors are the building blocks of our electronic technologies. 
They are used as fast electric current switches in every digital electronic device~\cite{MOSFET1,MOSFET2}. 
The application of transistors has been extended to atomic systems~\cite{AtomT1,AtomT2,AtomT3,AtomT4}, photon circuits~\cite{PhotonT1,PhotonT2}, and spintronics~\cite{spintronics1,spintronics2,MagnonT}.  
In spintronics, spin transistors have been developed as a controllable switch for transferring classical information encoded in spin degrees of freedom~\cite{Datta1990,Datta2018}.  
For miniaturizing spintronics as well as for going beyond classical computation, a new generation of transistors capable of acting at the single electron level would be highly desirable. 
Can the newly developed spin quantum simulators~\cite{quantumcomputation1,quantumcomputation2}, such as the Quantum dot arrays~\cite{QD1,QD2,QD3,QD4,QD5,QD6,QD7}, dopants in silicon~\cite{Dop1,Dop2,Dop3,Dop4,Dop5,Dop6,Dop7}, and molecular magnets~\cite{MolecularMagnet1,MolecularMagnet2,MolecularMagnet3,MolecularMagnet4}, be the arena to develop such transistors?

So far, Quantum Spin Transistors (QSTs) have been proposed for Heisenberg spin chains~\cite{QST1,QST2}, superconducting devices~\cite{QST3,QST4,QST5,QST10}, bosonic quantum oscillators~\cite{QST6} and symmetry protected many-body systems with adiabatic passage~\cite{QST7,QST8,QST9}. 
In these proposals,  the switching of the information flow in QSTs is controlled by either a magnetic~\cite{QST1,QST11,QST3,QST6,QST7,QST8} or a periodic driving~\cite{QST2} field. 
This makes the operation very slow and hard to realize. 
An open question is whether one can design a QST that is controlled  simply by external voltages, but still enables operations at the level of single electron spins.

Quantum state transfer in spin chains has been the subject of extensive research in the last decade~\cite{ST1,ST2,ST3,ST4}. Engineered chains~\cite{PST1,PST2,PST3,PST4,PST5,PST6,PST7}, quantum state routers~\cite{QR1,QR2}, arbitrary perfect state transfer methods~\cite{APST1,APST2}, induced direct interaction with perturbative methods~\cite{PerST1,PerST2,PerST3,PerST4}, many-qubit state transfer~\cite{MST1,MST2,MST3} and simultaneous quantum communication between multiple users~\cite{SST1,SST2} are just a few among many proposals for achieving high fidelity quantum state transfer in spin chains. The spin-only nature of these proposals implies that any QST based on these ideas will inevitably require a magnetic field for switching~\cite{QST1,QST2,QST3,QST4,QST5,QST6,QST7,QST8,
QST9,QST10}. In order to avoid this obstacle, one solution is to use charged particles which can be controlled by fast electric fields.

Here, we put forward a proposal for QSTs for itinerant electrons in a Fermi Hubbard lattice~\cite{bookHubbard,simulationHubbarddopant,
simulationHubbardot1,simulationHubbardot2,simulationHubbarsuperconduc,
simulationHubbarultracoldatom1,
simulationHubbarultracoldatom2,simulationHubbarion1,
simulationHubbarion2}. The charged configuration of the electrons, controlled by electric voltages in the gate port, can switch the information flow between the electrons trapped in the source and the drain of the transistor. While the transistor operates at the single electron level, its operation time is fast and switching can even happen within a fraction of this time. This allows for several subsequent operations of the transistor within the coherence time of the device. 
The robustness of the proposed QST against different types of noise is investigated and the obtained results based on realistic parameters demonstrate the capability of the QST to operate with high-fidelity in a realistic noisy setting. 
Therefor, it provides a new utilization for quantum dot arrays. \\

%

\begin{figure}[t!]
\includegraphics[width=\linewidth]{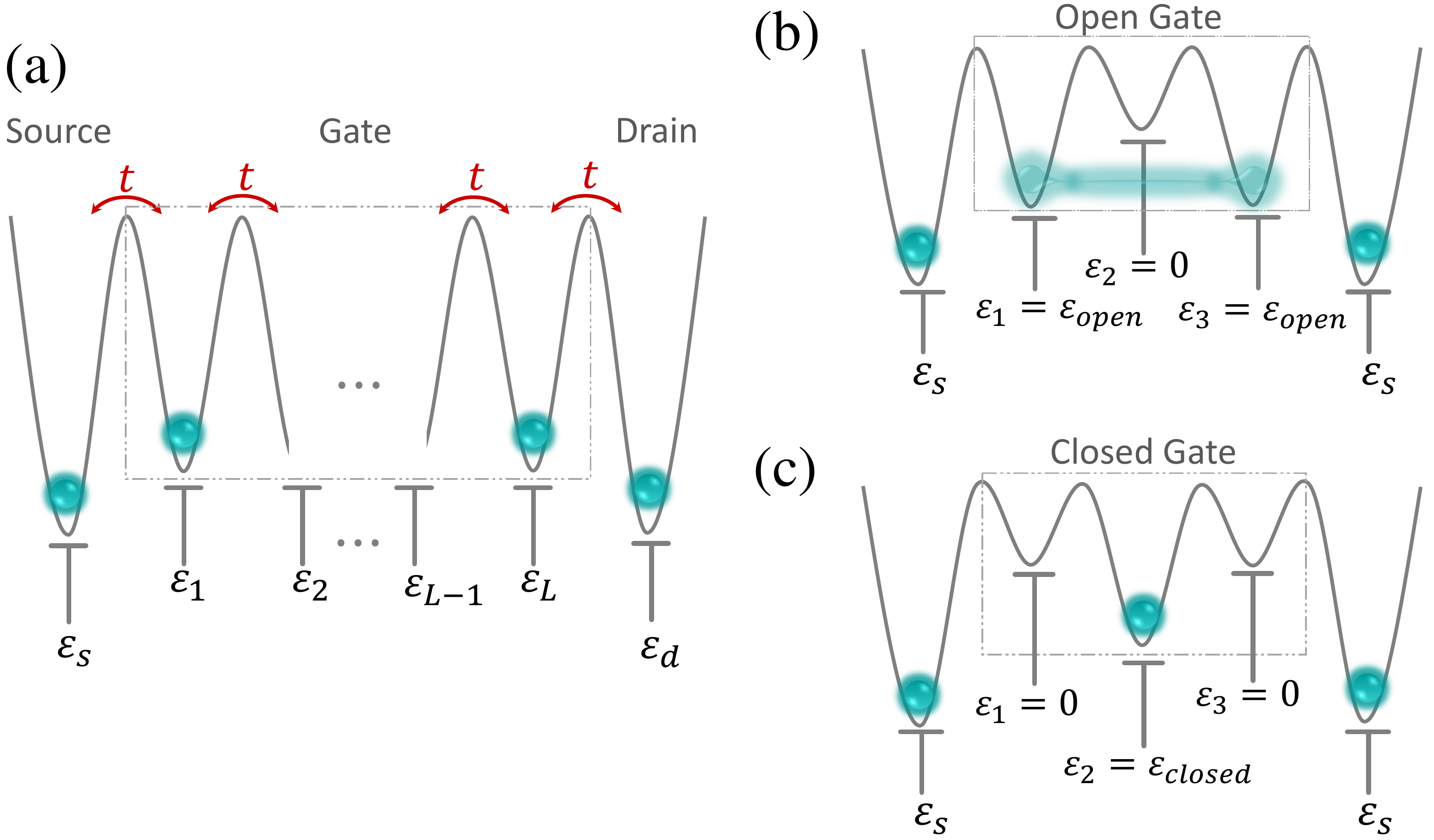}
\caption{(a) Schematic of the QST. In a 1D lattice,  
two electrons are extremely localized at source and drain sites and $n$ electrons hopping among $L$ sites of a central gate as quantum channel. (b) and (c) diagrams show how setting voltages can control the configurations of electron for opening and closing the gate in the shortest possible transistor with size $L{=}3$ filled by $n{=}1$ electron.}\label{fig:Schematic}
\end{figure}

\section{Model}
 We consider two extremely localized electrons as the source (i.e. input) and the drain (i.e. output) ports and $n$ electrons hopping among $L$ sites of a 1D lattice as the gate (i.e. channel). The transistor is shown schematically in Fig~\ref{fig:Schematic}(a). 
The dynamics of the overall system is ruled by an extended Fermi-Hubbard model $H_{tot}{=}H_{gate}{+}H_{s}{+}H_{d}{+}H_{I}$, in which we define    
\begin{eqnarray}\label{eq:Hubbard_Hamiltonian}
H_{gate}&=&-\sum_{k=1}^{L-1}\sum_{\sigma=\uparrow,\downarrow}t_{k,k+1}(c_{k,\sigma}^{\dagger} c_{k+1,\sigma} + h.c. )\cr &+& \sum_{k=1}^{L-1}V_{k,k+1} n_{k}n_{k+1}
+ \sum _{k=1}^{L}(U_{k} n_{k}^{\uparrow} n_{k}^{\downarrow} - \varepsilon_{k}n_{k}),\cr \cr
H_{s}&=&U_{s} n_{s}^{\uparrow} n_{s}^{\downarrow}
-\varepsilon_{s}n_{s},  \; \;\ \; \; \; \;\ \; \; \; \; \;\ \; \;
H_{d}=U_{d} n_{d}^{\uparrow} n_{d}^{\downarrow}
-\varepsilon_{d}n_{d}, \cr \cr
H_{I}&=&-\sum_{\sigma=\uparrow,\downarrow}(t_{s,1} c_{s,\sigma}^{\dagger} c_{1,\sigma} + t_{L,d} c_{L,\sigma}^{\dagger} c_{d,\sigma} + h.c. )\cr 
&+&  V_{s,1}n_{s}n_{1}+V_{L,d}n_{L}n_{d},
\end{eqnarray}
where, the subscript $s$ ($d$) represents the source (drain) qubit, and $H_{I}$ describes the interactions of the electrons in the source and the drain with the electrons in the gate.
In Eq.~(\ref{eq:Hubbard_Hamiltonian}),
$c_{k,\sigma}$ ($c_{k,\sigma}^{\dagger}$) is the annihilation (creation) fermionic operator
for an electron at site $k{\in}\{s,1,\ldots,L,d\}$ with spin $\sigma$, number operator $n_{k}{=}\sum_{\sigma=\uparrow ,\downarrow}n_{k}^{\sigma}$ with $n_{k}^{\sigma}{=}c_{k,\sigma}^{\dagger} c_{k,\sigma}$ counts the total number of electrons this site, $t_{k,k+1}$ is the tunneling between neighboring sites, $U_{k}$ is the on-site repulsion energy, $V_{k,k+1}$ is the Coulomb interaction between adjacent sites, and $\varepsilon_{k}$ is the local chemical potential at site $k$, which can be electrostatically controlled. 
We impose mirror symmetry (MS) on the chemical potential landscape $\vec{\varepsilon}{=}{[\varepsilon_{s},\varepsilon_{1},\ldots,\varepsilon_{L},\varepsilon_{d}]}$ such that  $\varepsilon_{s}{=}\varepsilon_{d}$ and $\varepsilon_{k}{=}\varepsilon_{L-k+1}$ for $k{\in}\{1,2,\ldots,\lceil{L/2}\rceil\}$. 
In the following sections we first take all the Hamiltonian parameters uniform i.e.,  $t_{k,k+1}{=}t$, $U_{k}{=}U$ and $V_{k,k+1}{=}V$ for all $k{\in}\{s,1,\ldots,L,d\}$ and also consider $U/t{=}50$ and $V/t{=}1$ which are matched with experimental realizations in quantum dot platforms~\cite{simulationHubbardot1,QD1}.
Next we investigate the role of non uniformity of the parameters in the performance of our protocol. 
All the dynamics are restricted to be within the time interval $\tau{\in}[0,500]{/}t$.

Note that for electron spin qubits in Si/SiGe quantum dots, one of the main concerns has been the valley degree of freedom, i.e. the degeneracy in the Si conduction band minima, with focus on the valley-orbit coupling induced by the interface disorder in material and its associated effect. 
Generally, valley orbit coupling $\Delta$ and the variations of valley phase across neighboring quantum dots can affect both the speed and the quality of coherent transport of electrons~\cite{valley1}.
Nevertheless, by setting the tunnel coupling $t$ and valley orbit coupling $\Delta$ more than Zeeman energies, the valley mixing can be completely suppressed~\cite{valley2}. By performing in this regime, the ground state manifold of the system is a spin doublet state with the same valley which does not overlap with the different valley (excited) states. In our protocol, thanks to the absence of magnetic field, this condition is satisfied and the valley degree of freedom is conserved for the ground state of the system.
\\

\section{Transistor modes}
 The spin transistor has two modes of operations determined entirely by the gate charge configuration. (i) open mode in which the gate allows for the flow of information between the source and the drain;  and (ii) closed mode in which the charge configuration of the gate blocks the exchange of information between the source and the drain. 
For these modes the sates of the gate are denoted by $\rho_{open}$ and $\rho_{closed}$, respectively.
Therefore, the operation of the spin transistor can be interpreted as a quantum Fredkin  gate~\cite{FredkinG1,FredkinG2,FredkinG3,QST5}, which is a controlled-swap gate and performs the state swapping between the source and the drain qubits, if the gate is open.

Initially, the electron in the source site has a specific spin state $\vert\psi_{s}\rangle{=}\alpha \vert{\uparrow_{s}}\rangle{+}\beta \vert{\downarrow_{s}}\rangle $, where $\vert{\uparrow_{s}}\rangle$ ($\vert{\downarrow_{s}}\rangle$)  represents one electron with spin up (down) at site $s$. The localized electron at the drain site is assumed to be in the maximally mixed state $\mathcal{I}_{d}{=}(\vert{\downarrow_{d}}\rangle\langle{\downarrow_{d}}\vert {+} \vert{\uparrow_{d}}\rangle\langle{\uparrow_{d}}\vert)/2$ which indicates no control over the spin state of the electron. 
The gate is initialized in the ground state of $H_{gate}$, denoted by $\rho_{mode}{\in}\{\rho_{open},\rho_{closed}\}$.  
The initial state of the whole system is thus described by $\rho(0){=}\vert\psi_{s}\rangle\langle \psi_{s}\vert {\otimes}\rho_{mode} {\otimes}\mathcal{I}_{d}$. Then, the system evolves under the action of the total Hamiltonian $H_{tot}$ as $\rho(\tau){=}U_{\tau}\rho(0)U_{\tau}^{\dagger}$, where $U_{\tau}{=}e^{-iH_{tot}\tau}$. Ideally, in the open mode the target state is $\rho_{open}^{target}{=} \mathcal{I}_{s} {\otimes}\rho_{open} {\otimes} \vert\psi_{d}\rangle\langle \psi_{d}\vert$, which is obtained by swapping the quantum states of the source and the drain in the initial state. On the other hand, in the closed mode, the system should remain unchanged and thus the target state is $\rho_{closed}^{target}{=}\vert\psi_{s}\rangle\langle \psi_{s}\vert {\otimes}\rho_{closed} {\otimes}\mathcal{I}_{d}$.  To quantify the performance of the transistor at each mode (i.e., either open or closed), we compute the fidelity $\mathcal{F}_{mode}(\tau){=}F(\rho(\tau),\rho_{mode}^{target})$. 
By taking the average over all possible input states, one gets 
the input-independent quantity $\overline{\mathcal{F}}_{mode}(\tau){=}\int \mathcal{F}_{mode}(\tau) d\Omega$, in which $d\Omega$ is the Haar measure over the Bloch sphere. While $\overline{\mathcal{F}}_{open}(\tau)$ takes its maximum at time $\tau{=}\tau^{opt}$, which will be the operation time of the transistor,  $\overline{\mathcal{F}}_{closed}(\tau)$ should remain closed to $1$ at all times.
  \\


\begin{figure}[t!]
\includegraphics[width=\linewidth]{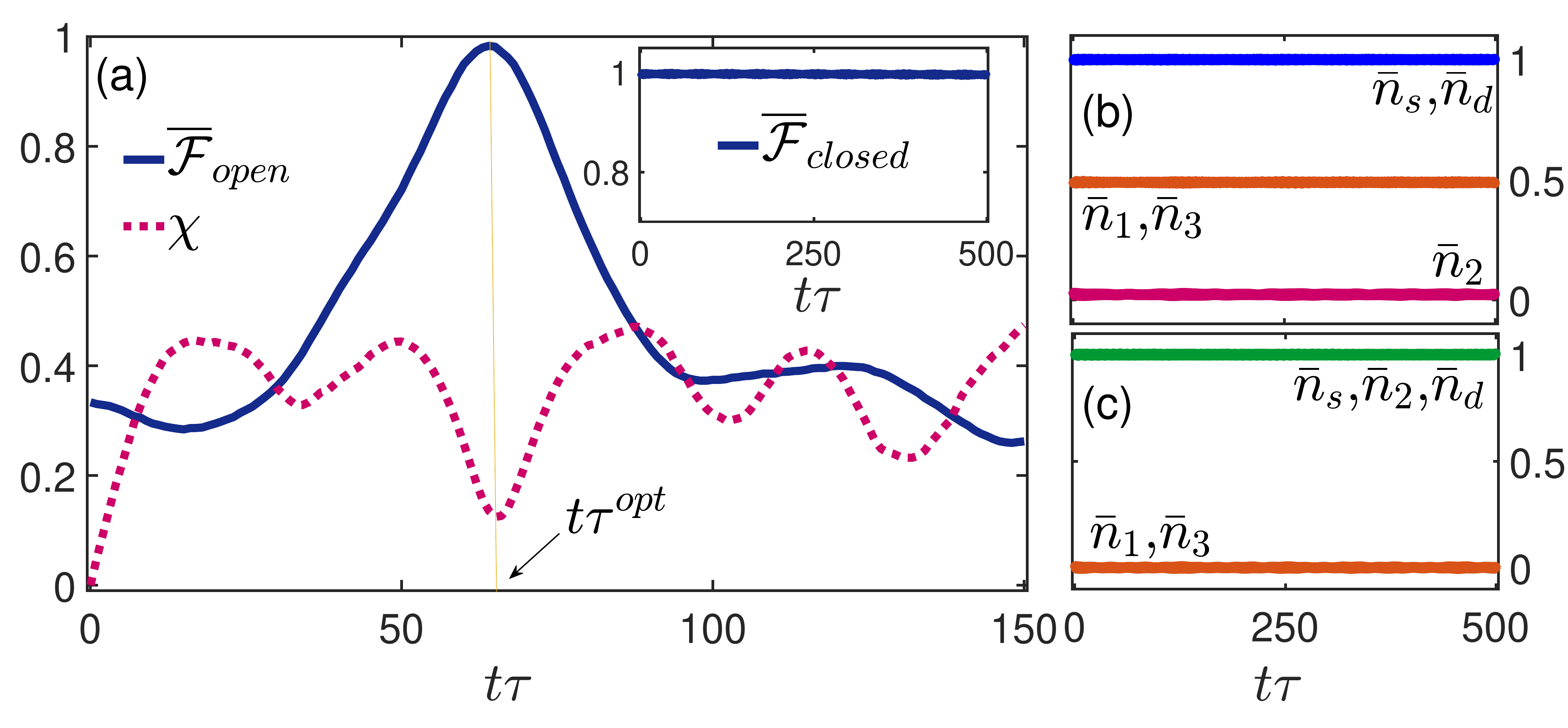}
\caption{ (a) Dynamics of the average fidelity and logarithmic negativity $\chi$ in the shortest ($L{=}3$ and $n{=}1$) transistor with open gate. The inset shows the performance of the transistor with closed gate. 
The charge occupancy diagrams in the shortest transistor  with open (b) and closed (c) gate.
In all plots, the potential landscape are tuned in optimal values $\varepsilon_{open}^{opt}{=}10t$, $\varepsilon_{closed}^{opt}{=}20t$, and $\varepsilon_{s,d}^{opt}{=}39t$ for both modes.}\label{fig:F-t(L3,N1)}
\end{figure}

\subsection{Optimizing charge configuration}
The key point in the performance of the transistor is to tune the chemical potential $\vec{\varepsilon}$ to: (i) keep the electrons in the source and the drain sites localized at all times; and (ii) get proper quantum states of the gate, namely $\rho_{open}$ and $\rho_{closed}$, in order to maximize the fidelity $\overline{\mathcal{F}}_{mode}(\tau^{opt})$ for both modes. 
For the open mode, we push the electron inside the gate to be delocalized at the edges by choosing the electron configuration as $\varepsilon_{k}{=}\vert\lceil{L{/}2}\rceil{-}k \vert\varepsilon_{open}$, see Fig.~\ref{fig:Schematic}(b). 
For the closed mode, the electron is localized at the middle site by setting $\varepsilon_{k}{=}(k{-}1)\varepsilon_{closed}$, see Fig.~\ref{fig:Schematic}(c). 
Hence, $\varepsilon_{open}$ and $\varepsilon_{closed}$ control the entire chemical potentials inside the gate. This is in particular useful for dopant-based systems in which single site control may not be available.  At each mode, $\varepsilon_{mode}$ together with the parameter $\varepsilon_s{=}\varepsilon_d$ should be optimized so that the above two conditions are satisfied for an optimal operation time $\tau^{opt} {\in} [0,500]/t$. 

\begin{table*}[t!]
{
\renewcommand{\arraystretch}{1.55}
\setlength{\tabcolsep}{1.9pt}
\begin{tabular}{|c|c|c|c|c|c|c|c|c|c|c|c|c|c|c|c|c|c|c|c|}
\hline
$L$ & \multicolumn{2}{c|}{3} & \multicolumn{4}{c|}{4}        & \multicolumn{5}{c|}{5}                & \multicolumn{6}{c|}{6}                        & 7     & 8     \\ \hline \hline
$n$ & 1          & 2         & 1     & 2     & 3     & 4     & 1     & 2     & 3     & 4     & 5     & 1     & 2     & 3     & 4     & 5     & 6     & 1     & 1     \\ \hline
$\overline{\mathcal{F}}_{open}(\tau^{opt})$ & 0.983      & 0.974     & 0.972 & 0.871 & 0.970 & 0.982 & 0.959 & 0.651 & 0.961 & 0.723 & 0.821 & 0.969 & 0.414 & 0.930 & 0.764 & 0.791 & 0.887 & 0.925 & 0.921 \\ \hline
$t\tau^{opt}$ & 64         & 459       & 81    & 286   & 107   & 255   & 129   & 484   & 409   & 270   & 434   & 153   & 492   & 493   & 346   & 455   & 424   & 184   & 270   \\ \hline
$\varepsilon_{s,d}^{opt}/t$  & 39         & 74        & 30   & 25  & 85  & 23  & 37 & 36 & 93 & 31 & 99 & 37 & 39 & 85 & 32 & 75 & 49  & 35 & 39 \\ \hline
$\varepsilon_{open}^{opt}/t$ & 10         & 1         & 4    & 1   & 55  & 35  & 3  & 1  & 39 & 5  & 1  & 2  & 1  & 4  & 1  & 1  & 23  & 1  & 1  \\ \hline
$\varepsilon_{close}^{opt}/t{\geq}$ & 20         & 54        & 14   & 10  & 54  & 64  & 3  & 6  & 9  & 50 & 40 & 2  & 3  & 8  & 9  & 31 & 3  & 1  & 2  \\ \hline
\end{tabular}
\caption{ The achievable $\overline{\mathcal{F}}_{open}(\tau^{opt})$ at different size $L{=}3,\ldots,8$ with various number of electrons $1{\leq}n{\leq}L$ which is obtained at optimal time $\tau^{opt}$ and optimal potential $\varepsilon_{s}^{opt}$ and $\varepsilon_{open}^{opt}$~\cite{Note}. }\label{table}
}
\end{table*}

For the sake of simplicity, we start with the shortest possible gate consisting of $L{=}3$ sites and $n{=}1$ electron, for which we can provide analytic solution. Longer gates and larger filling factors will be discussed later.
In general, due to the absence of magnetic field, the ground state of the gate is spin degenerate with the same charge configurations. Therefore, we consider the state of the gate to be an equal mixture of all those spin-degenerate ground states.
By choosing $\varepsilon_{open}$ and $\varepsilon_{closed}{\gg}{2\sqrt{2}}t$ (see the Appendix for details), the ground state of  $H_{gate}$ in both open (i.e. $\vert O_{\sigma}\rangle$) and closed (i.e. $\vert C_{\sigma}\rangle$) modes is doubly degenerate and is given by 
\begin{eqnarray}\label{eq:ground states}
\vert O_{\sigma}\rangle = (\vert{\sigma,0,0}\rangle{+}\vert{0,0,\sigma}\rangle)/\sqrt{2} \;\;\;\mathrm{and}\;\;\;
\vert C_{\sigma}\rangle = \vert{0,\sigma,0}\rangle
\end{eqnarray}
where, $|0\rangle$ represents an empty site and $|\sigma\rangle$ stands for a site containing one electron with spin $\sigma{=}\uparrow , \downarrow$. Therefore, one gets $\rho_{open}{=}1/2\sum_{\sigma} \vert{O_\sigma}\rangle\langle O_{\sigma}\vert$ and $\rho_{closed}{=}1/2\sum_{\sigma} \vert{C_\sigma}\rangle\langle C_{\sigma}\vert$. 
By exploiting the brute force optimization, we obtain  $\varepsilon_{open}^{opt}{=}10t$ and $\varepsilon_{s,d}^{opt}{=}39t$ for the open mode. 
For the closed mode, we keep $\varepsilon_{s,d}^{opt}$ fixed as in the open mode and optimize the average fidelity with respect to $\varepsilon_{closed}$ which results in  $\varepsilon_{closed}^{opt}{=}20t$. 
Using these optimal parameters,  we plot $\overline{\mathcal{F}}_{open}$ as a function of time in Fig.~\ref{fig:F-t(L3,N1)}(a) which shows that the fidelity gradually increases and at a specific time $\tau^{opt}{=}64{/}t$ peaks to its highest value $\overline{\mathcal{F}}_{open}(\tau^{opt}){=}0.983$, resulting in nearly perfect information transmission.
The inset of Fig.~\ref{fig:F-t(L3,N1)}(a) depicts $\overline{\mathcal{F}}_{closed}$ versus time and shows that, in closed mode, the fidelity remains very close to $1$ throughout the evolution.
To fully understand the underlying mechanism of our QST, one has to discriminate the charge and spin degrees of freedom. As shown in Fig.~\ref{fig:F-t(L3,N1)}(b), for both operational modes of the transistor, the charge occupancy $\bar{n}_{k}{=}Tr(n_{k}\rho(\tau))$ of each site $k{\in}\{s,1,\ldots,L,d\}$ presents very stable behavior during the evolution.  
By tuning  $\varepsilon_s{=}\varepsilon_d$ to be off resonant with the other chemical potentials in the gate, one can trap the electrons in the source and the drain.
However, thanks to the uniformity of the tunneling $t$, the spin flow from the source to gate is not energetically costly and thus spin information flows directly through the gate. 
This can be illustrated by the entanglement dynamics of the gate with the rest of the system, quantified by the logarithmic negativity $\chi(\tau){=}\log_{2}|\rho^{T_{g}}(\tau)|$ and plotted in Fig.~\ref{fig:F-t(L3,N1)}(a). Here, $T_{g}$ stands for partial transpose and $|\cdot|$ represents the trace norm. 
The generated entanglement during the dynamical evolution is a clear witness for the spin flow of information between the source and the drain through the electrons inside the gate.
Obviously, this mechanism is very  distinct from perturbative methods for quantum state transfer~\cite{PerST1,PerST2,PerST3} in which very few (usually two) eigenstates with bi-localized charges at the edges participate in the transmission dynamic.

In the same spirit, one can construct the QSTs with longer gates containing only one electron. 
From the practical point of view, increasing the distance between the source and the drain qubits helps to avoid crosstalk and achieve better local addressability and isolation. In addition, the Hubbard model can be an approximation for a gate made of a large quantum dot with multiple electrons.
By optimizing the chemical potential landscape $\vec{\varepsilon}$ (presented in TABLE~\ref{table}) one can  induce a high quality spin flow of information.
In TABLE~\ref{table}, for different gate length $L{=}3,\ldots,8$ with $n{=}1$ electron, the obtainable average fidelity in the open mode, the operation time $\tau^{opt}{\in}[0,500]/t$ and optimal values of the chemical potentials for both modes are reported.
Note that through out this paper, the chemical potentials are obtained by exploiting the brute force optimization over the interval $[0,100]$ and result in high fidelity operations of the transistors within the operational time interval $\tau{\in}[0,500]/t$.
In the case of $n{=}1$, by increasing $L$, the charge density in the gate leaks from the edges to the internal sites. This results in  slow decay of the average fidelity from $\overline{\mathcal{F}}_{open}(\tau^{opt}){=}0.983$ for $L{=}3$ to  , $\overline{\mathcal{F}}_{open}(\tau^{opt}){=}0.921$ for $L{=}8$. At the same time, the operation time $\tau^{opt}$ increases almost linearly by $L$.\\

\begin{figure}[t!]
\includegraphics[width=\linewidth]{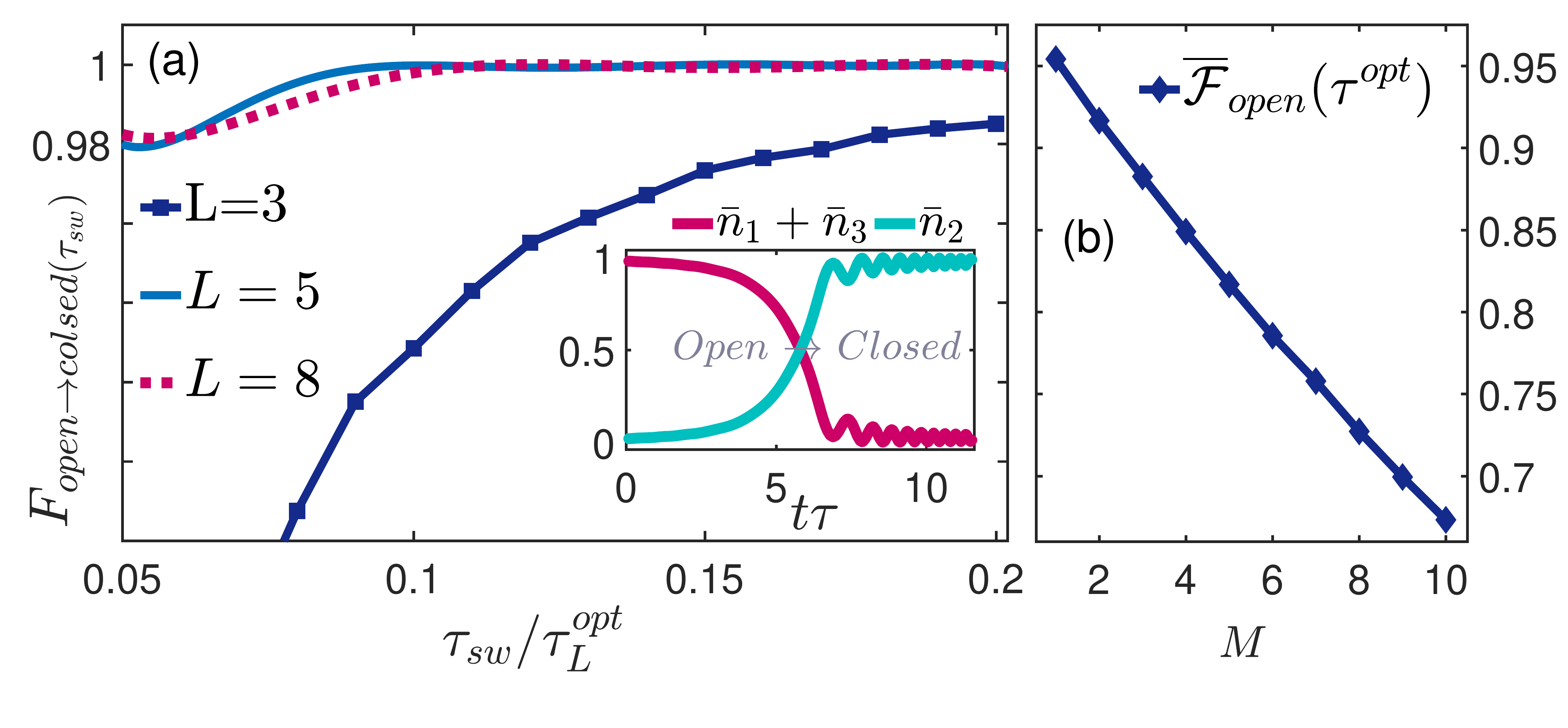}
\caption{ (a) Fidelity of transition $\rho_{open}$ into $\rho_{closed}$ at the end of the bias sweeping as a function of $\tau_{sw}$ which is a percentage of $\tau^{opt}_{L}$ for different gate length $L{=}3,5,8$ and $n{=}1$. Inset shows the transmission of the electron from the barriers to the middle of the gate by bias sweeping from $\varepsilon_k{=}|\lceil{L/2}\rceil{-}k|\varepsilon_{open}$ into the opposite bias $\varepsilon_k{=}(k{-}1)\varepsilon_{closed}$ as a function of time in the shortest transistor with $n{=}1$. (b) Functionality of the shortest transistor with $L{=}3$ and $n{=}1$ for $M$ subsequent opening and closing the gate without resetting. }\label{fig:open2close}
\end{figure}

\section{Switching the transistor}
In order to be practically useful, one has to be able to rapidly switch between the open and the closed modes. In fact, since the gate in both modes operates in its ground state, the transition can be implemented by adiabatically sweeping the bias chemical potentials to the desired  values over a switching time $\tau_{sw}$.
For transmission from the open to the closed mode,  
in the first half of the switching time, $0{\leq}\tau{\leq}\tau_{sw}{/}2$, the local chemical potential on the gate is swept to zero through  
$\varepsilon_k(\tau){=}(1{-}2\tau{/}\tau_{sw})|\lceil{L/2}\rceil{-}k|\varepsilon_{open}$.
Then the sweeping is followed by $\varepsilon_k(\tau){=}(2\tau{/}\tau_{sw}{-}1)(k{-}1)\varepsilon_{closed}$ in the second half of the switching time, $\tau_{sw}{/}2{<}\tau{\leq}\tau_{sw}$. 
The evolution of the gate is thus given by   
$V_{\tau}{=}\mathcal{T}e^{-i\int_{0}^{\tau}H_{gate}(\tau^{\prime})d\tau^{\prime}}$ with $\mathcal{T}$ being the time-ordering operator. 
The state of the gate evolves as $\rho_{gate}(\tau){=}V_{\tau}\rho_{open}V_{\tau}^{\dagger}$. For evaluating this transition, one can consider the fidelity $F_{open\rightarrow closed}(\tau_{sw}){=}F(\rho_{closed},\rho_{gate}(\tau_{sw}))$.
Similarly, the transition from the closed to the open mode can be carried out by adiabatically sweeping the chemical potentials from the closed to the open mode.
The quality can be quantified via transition fidelity $F_{closed\rightarrow open}(\tau_{sw}){=}F(\rho_{open},\rho_{gate}(\tau_{sw}))$. We search for the lowest $\tau_{sw}$ so that these transmission fidelities remain very high. 
In Fig.~\ref{fig:open2close}(a) we plot $F_{open{\rightarrow}closed}(\tau_{sw})$ as a function of switching time $\tau_{sw}$ for gate of sizes $L{=}3,5,8$ filled by $n{=}1$ electron. Interestingly, for a gate of size $L{=}3$ by choosing $\tau_{sw}/\tau^{opt}{\simeq} 0.18$ one can achieve a transmission fidelity of $F_{open{\rightarrow}closed}(\tau_{sw}){>}0.98$. This means that one can switch between the modes of the transistor within a fraction of its operation time. The inset of  Fig.~\ref{fig:open2close}(a) shows the charge movements between different sites of the gate. The initially delocalized electron between the edge sites becomes localized at the middle site after the switching time $\tau_{sw}$. 
The larger gates operate in the same way. For instance, for the gate size $L{=}8$ and  $n{=}1$ the switching time can be as small as  $\tau_{sw}/\tau^{opt}{=}0.05$ to reach the fidelity $F_{open{\rightarrow}closed}(\tau_{sw}){>}0.98$.
Interestingly, the switching from the closed to the open mode can be achieved within the same switching time $\tau_{sw}$.
It is desirable to see how the quality of the transistor operation is
affected by $M$ subsequent opening and closing the gate without resetting the transistor. For a transistor of gate size $L{=}3$ filled by $n{=}1$ electron, we  plot the obtainable fidelity $\overline{\mathcal{F}}_{open}$ as a function of $M$ in Fig.~\ref{fig:open2close}(b). Although, the quality of the transmission decays with subsequent uses, even after $M{=}10$ switching the fidelity still remains above the classical threshold $2{/}3$. \\

\section{Filling factor effect}  
Here, we investigate the role of increasing the number of electrons in the gate (namely, $n$) on the performance of the open mode.  
TABLE~\ref{table} presents the obtainable average fidelity $\overline{\mathcal{F}}_{open}(\tau^{opt})$, the optimal time $\tau^{opt}$ and corresponding optimal chemical potentials for various choices of $3{\leq}L{\leq}8$ and $1{\leq}n{\leq}L$. 
While the gates with odd $n$ have twofold degenerate ground state with total spin $S_{tot}{=}1/2$, the ground state for even $n$ electrons is a global spin-singlet with $S_{tot}{=}0$.
This results in different outcomes for even and odd $n$.  

\begin{figure}[t!]
\includegraphics[width=\linewidth]{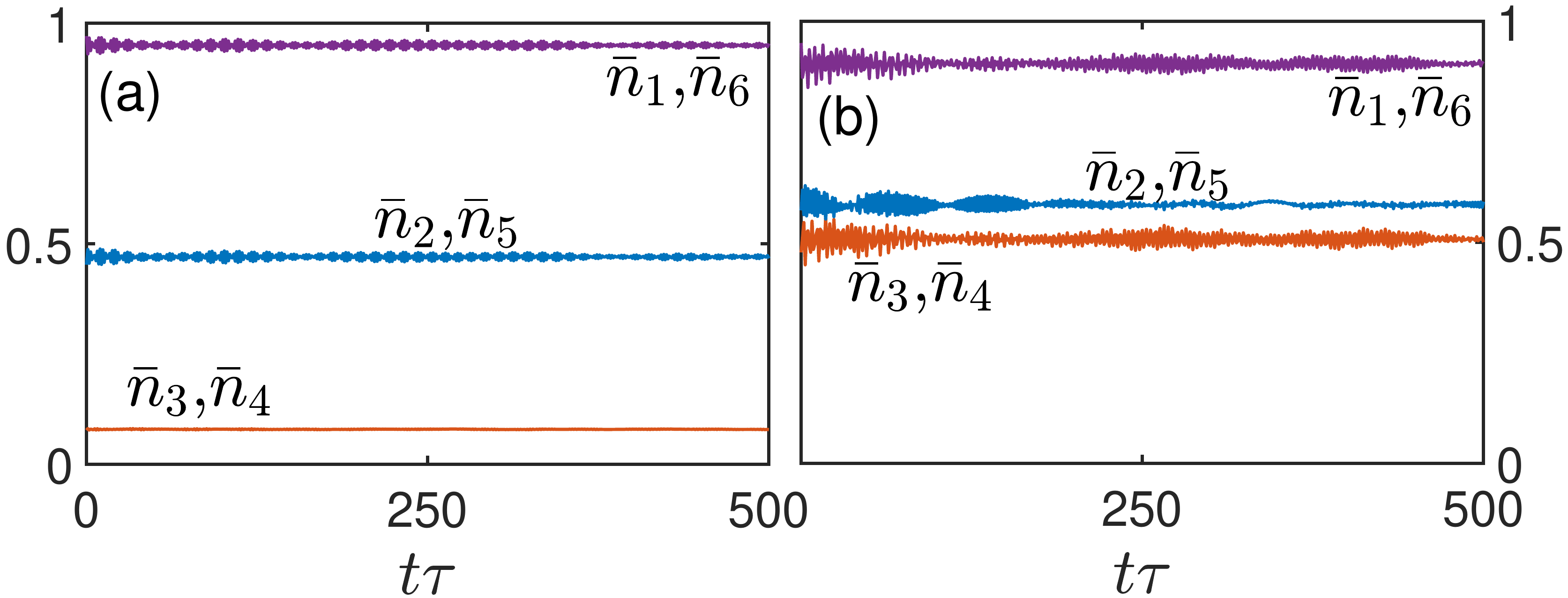}
\caption{Charge occupancies of a
gate of length $L{=}6$ filled by $n{=}3$ (a), and $n{=}4$ (b) electrons versus time. In all plots the chemical potential are tuned in their optimal values presented in TABLE~\ref{table}.}\label{fig:ns}
\end{figure}

Let's first focus on filling factors less than half,  $n{<}L$.
In the case of odd $n$, the $n{-}1$ electrons localize at the corner sites of the gate with $\bar{n}_k{\sim} 1$. 
Due to the mirror symmetry, the last electron becomes delocalized between the two sites adjacent to the occupied ones with $\bar{n}_k{\sim} 0.5$. The rest of the inner sites have small charge occupancy (i.e., $\bar{n}_k{\ll} 1$). 
This can be seen from Fig.~\ref{fig:ns}(a) which illustrates the charge occupancies in the evolution of a transistor of length $L{=}6$ which is filled by $n{=}3$ electrons.
Obviously, the sites with lowest energies, i.e., the corners, are fully occupied by $2$ electrons (one electron per site) and the remained electron tends to be delocalized on adjacent empty sites with upper energies, i.e., sites 2 and 5.
Based on the TABLE~\ref{table}, for a fixed $L$, increasing two electrons (keeping $n$ odd), keeps the average fidelity almost the same and increases the operation time.
One can explain this as the localized electrons at the corner sites create an effective barrier for information flow which slows the dynamics. 
In the case of even $n$, while $n{-}2$ electrons are localized in the corner sites (with $\bar{n}_k{\sim} 1$) the two remaining electrons become highly delocalized among all the internal sites (see Fig.~\ref{fig:ns}(b) for the dynamics of the charge occupancies in transistor of $L{=6}$ and $n{=}4$). 
For a fixed $L$, increasing two electrons (keeping $n$ even), enhances the average fidelity and shrinks the operation time significantly. 
In fact by increasing $n$, the two delocalized electrons disperse over inner sites, thus the wave function overlap between internal sites increases. 
This results in faster dynamics. 
Note that, in TABLE~\ref{table}, some of the cases of even $n$ show very low fidelity which is due to slow dynamics and the insufficiency of the considered time interval.

In the half filling scenario, $n{=}L$, the dynamics of the system in the open mode can be reduced to the extensively studied quantum state transfer in spin chains~\cite{ST1,ST2,ST3,ST4}. 
For this case, in compare to lower filling factors, the dynamics is slower and the average fidelity may take lower values. 
This shows the advantage of the itinerant electrons over spin chain state transfer.

Note that, for all considered $3{\leq}L{\leq}8$ and $1{\leq}n{\leq}L$, the closed mode can be perfectly obtained merely by tuning the chemical potential $\vec{\varepsilon}$ and localizing the electrons in the central sites of the gate. In TABLE~\ref{table}, the corresponding optimal chemical potential  $\varepsilon_{closed}^{opt}$  for closed mode is presented for various $3{\leq} L{\leq} 8$, and $1{\leq} n{\leq} L$.
In the case of $L{=}3$ and $n{=}3$, there is no configuration that keeps all of the electrons far away the source and the drain qubits, so the closed mode can not be faithfully captured and is omitted.

\section{Operation under realistic conditions}
In order to capture all relevant effects of implementing our proposal based on different types of quantum platforms, we first consider the effects of temperature, disorder in Hamiltonian parameters, and electric charge noise, as common concerns in all physical platforms. 
Next we devote the largest attention to gate-controlled quantum dot arrays~\cite{QD1,QD2,QD16,QD17} as the most suitable physical platforms to realize our proposed QST. Quantum dots naturally realize the Hubbard Hamiltonian~\cite{simulationHubbardot1,
simulationHubbardot2,simulationHubbarultracoldatom1,
simulationHubbarultracoldatom2} and the coherence time of such systems has reached to $T_{2}{\sim}100 \mu s$~\cite{QD5,QD8,QD9,QD15,QD10,QD7}, which can be further enhanced through environmental charge noise suppression~\cite{QD11,QD12} or material engineering~\cite{QD13}. 
Quantum dots take a variety of forms with specific mechanisms that lead to spin flip.   
In GaAs quantum dots this mechanism is the spin-orbit coupling (SOC) and in the following the effect of this source of noise on the performance of the transistor is considered.

\subsection{Thermal Effect}
In the case of nonzero temperature, the gate is initialized in a thermal state $\rho_{gate}{=}e^{-H_{gate}/TK_{B}}{/}tr[e^{-H_{gate}/TK_{B}}]$ rather than a mixture of spin degenerate ground states, where $T$ and $K_{B}$ are the temperature and Boltzmann constant, respectively. If the thermalization time is longer than the performance of the transistor then the dynamics remain unitary. In Fig.~\ref{fig:realistic}(a), the effect of the temperature on the functionality of the open and closed gates with $L{=}3$ and $n{=}1$ is reported. 
While, by increasing the temperature  $\overline{\mathcal{F}}_{open}(\tau^{opt})$ decays slowly, the closed transistor demonstrates very robust behavior as long as $TK_{B}{/}t{\ll}|\varepsilon_{1}{-}\varepsilon_{2}|$.
In a typical quantum dot platform with $t{=}0.02$ meV,  one can get $\overline{\mathcal{F}}_{open}(\tau^{opt}){>}0.9$ for temperature close to $60$ mK.

\subsection{Disorders}
\begin{figure}[t!]
\includegraphics[width=\linewidth]{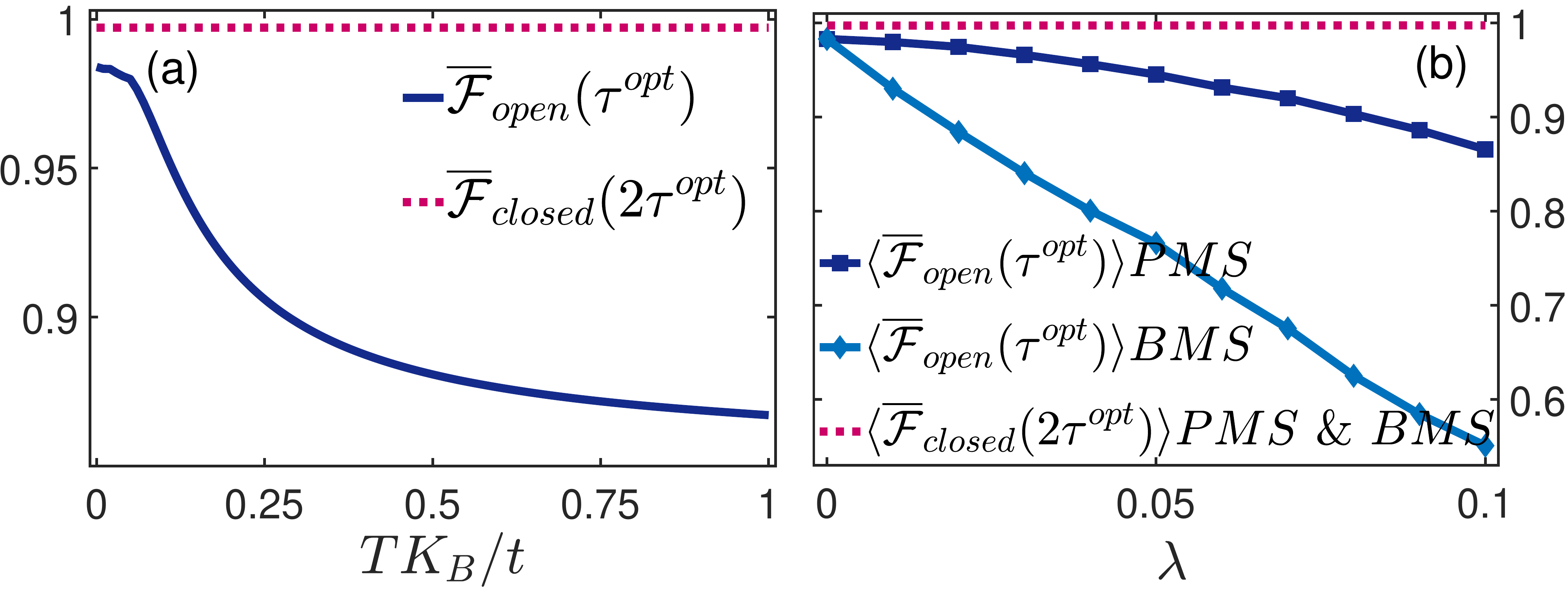}
\includegraphics[width=\linewidth]{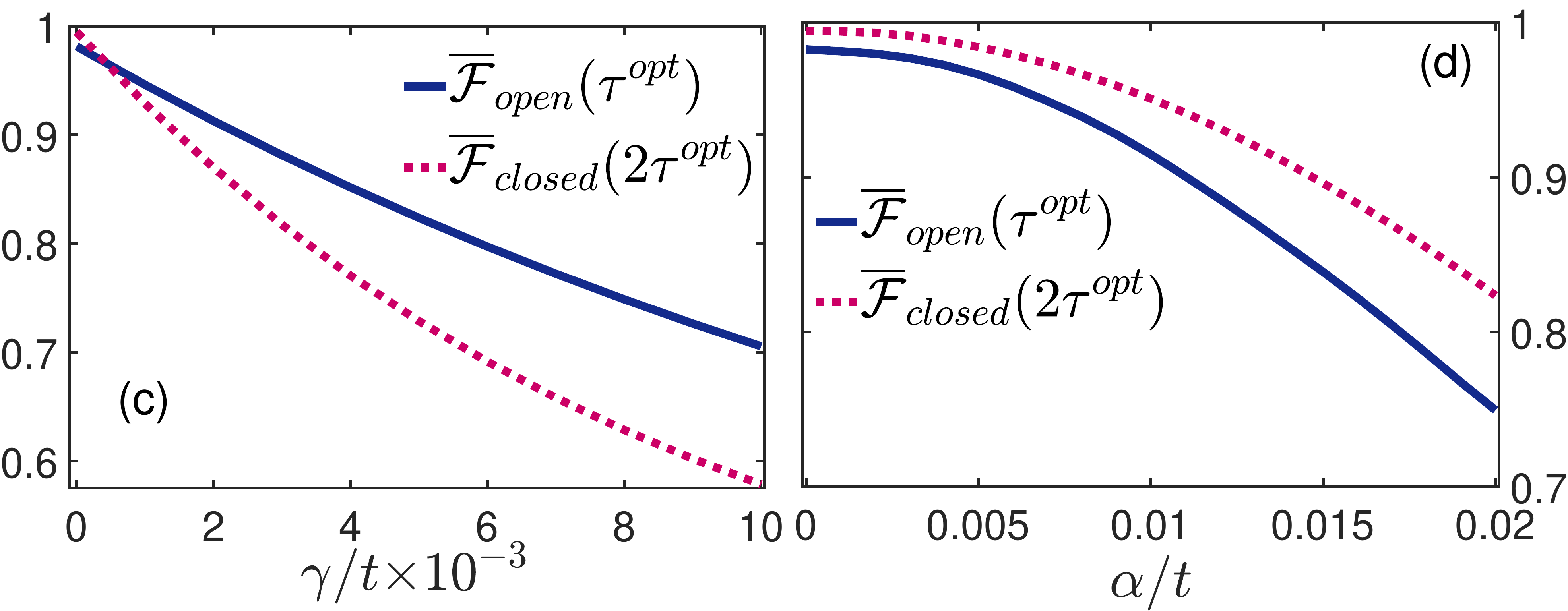}
\caption{Functionality of the shortest transistor filled by $n{=}1$ electron with open and closed gates as a function of the temperature $TK_{B}$ (a), disorder strength $\lambda$ (b), decoherence rate $\gamma$ (c), and SOC strength $\alpha$ (d).  
In these plots the parameters are tuned as $\tau^{opt}{=}64t$, 
$\varepsilon_{s,d}^{opt}{=}39t$, $\varepsilon_{open}^{opt}{=}10t$ and 
$\varepsilon_{closed}^{opt}{=}20t$. }\label{fig:realistic}
\end{figure}

The key assumptions in the performance of the transistor are the homogeneity of the Hamiltonian parameters and mirror symmetry (MS) on the applied chemical potentials. 
However, fabrication disorder in any experimental setup is inevitable. As a consequence the Hamiltonian parameters may subject to random variations and stay nonuniform. 
For analyzing the effects related to randomness of tunneling, one-site repulsion energy and Coulomb interaction we consider the parameters in the Hamiltonians of Eq.~(\ref{eq:Hubbard_Hamiltonian}) as  $t_{k,k+1}{=}t(1{+}\Lambda_{t})$, $U_{k}{=}U(1{+}\Lambda_{U})$, and $V_{k,k{+}1}{=}V(1{+}\Lambda_{V})$ for $k{\in}\{s,1,\ldots,L,d\}$.
Moreover, to investigate the effect of disorder in applied chemical potentials we first replace $\varepsilon_{s,d}$ with $\varepsilon_{s,d}^{opt}(1{+}\Lambda_{s})$ and also $\varepsilon_{k}$ with $\varepsilon_{k}^{opt}(1{+}\Lambda_{k})$ in a way that $\varepsilon_{k}^{opt}(1{+}\Lambda_{k}){=}\varepsilon_{L-k+1}^{opt}(1+\Lambda_{k})$ for $k{\in}\{1,2,\ldots,\lceil{L/2}\rceil \}$. 
By this assumption the MS condition is preserved.
Next we relax this assumption and consider $\varepsilon_{k}^{opt}(1{+}\Lambda_{k})$ as the local chemical potential on all sites $k{\in}\{s,1,\ldots,L,d\}$ to break down the MS. 
Although, in real experiments, the strength of disorder for each parameter is different, here, for the sake of simplicity, we assume that all $\Lambda_{i}$ for $i{\in}\{t,U,V,s,k\}$ are sampled from a uniform distribution with 
$\Lambda_{i}{\in}[{-}\lambda,\lambda]$.
The advantage of this choice is that with controlling $\lambda$ one can investigate the effect of disorder in all the parameters. 
By generating 500 different random Hamiltonians $H$, according to
this distribution, for each value of $\lambda$, we calculate $\overline{\mathcal{F}}_{open}(\tau^{opt})$ and $\overline{\mathcal{F}}_{closed}(2\tau^{opt})$ for open and closed modes.
Averaging over all
these random realizations one gets $\langle\overline{\mathcal{F}}_{open}(\tau^{opt})\rangle$ and $\langle\overline{\mathcal{F}}_{closed}(2\tau^{opt})\rangle$ to quantify the quality of the transistor's modes.
The performance of shortest transistor with $L{=3}$ and $n{=}1$ as a function of $\lambda$ at both modes while MS is preserved (PMS) and broken (BMS) is plotted in Fig.~\ref{fig:realistic}(b).
While by increasing the strength of the disorder, $\lambda$, the quality of the transmission in open mode decreases, closed mode remains totally stable. 
Regarding the configuration of the electron inside the gate in the closed mode, this robust behavior is expected. 
In this case, the strength of the disorder is not enough for delocalizing  electron in the middle of the gate and, hence, transferring the information.  
In the presence of MS, the protocol
shows more robust behavior and even for a disorder with strength $\lambda{<}0.08$, one gets  $\langle\overline{\mathcal{F}}_{open}(\tau^{opt})\rangle{>}0.9$.
Clearly, breaking the SM results in sharp reduction of transmission's quality and $\langle\overline{\mathcal{F}}_{open}(\tau^{opt})\rangle{>}0.9$ can be only obtained for $\lambda{<}0.018$.

\subsection{Electric charge noise}
Another source of imperfections which can be more serious is electric charge noise.     
As this noise imposes a challenge for most operations in  electron spin qubits, a thorough characterization of it is essential.
A possible explanation for this noise is that each eigenstate of the system, i.e. $\{|E_{i}\rangle\}$, experiences different interaction with the environment resulting in a Markovian dynamics described by a Linbladian master equation 
\begin{equation}\label{eq:Master}
\dot{\rho}(\tau)=-i[H,\rho(\tau)]+\gamma \sum_{i}(\Gamma_{i}\rho(\tau)\Gamma_{i}^{\dagger}-\dfrac{1}{2}\{\Gamma_{i}^{\dagger}\Gamma_{i},\rho(\tau)\}),
\end{equation}
where $\gamma$ is the strength of the noise and the Lindblad operators are given as  $\Gamma_{i}{=}\vert{E_{i}}\rangle\langle{E_{i}}\vert$.   
The functionality of the shortest transistor with $L{=}3$ filled by one electron as a function of $\gamma$  for both operational modes is presented in Fig.~\ref{fig:realistic}(c).
As the noise strength enhances the fidelities decays to lower values. Nonetheless, for $\gamma{<}0.002t$, both average fidelities exceed $0.9$ showing that our transistor can tolerate resonantly strong charge fluctuations.

\subsection{Spin-orbit coupling}
In some of the quantum dot platforms (e.g. GaAs and InAs) one of the main concerns has been SOC, a mechanism that leads to spin flipping. 
For 1D Fermi-Hubbard model that we have here, the SOC Hamiltonian  accounting for the well known Rashba and Dresselhaus effects is given by~\cite{QD14}
\begin{equation}
H_{soc} = -t\zeta (c_{s,\uparrow}^{\dagger} c_{1,\downarrow} +  \sum_{k=1}^{L-1}  c_{k,\uparrow}^{\dagger} c_{k+1,\downarrow} + c_{L,\uparrow}^{\dagger} c_{d,\downarrow}) + h.c.,
\end{equation}\label{Eq:spin-orbit}
where $\zeta{=}\alpha{-}i\beta$ and $\alpha$ and $\beta$ are the Rashba and the Dresselhaus parameters for typical quantum dots and depends on the internal properties and spatial gaps between the quantum dots.
In this case the extended Hubbard model $H_{tot}^{\prime}{=}H_{tot}{+}H_{soc}$ rules the dynamics of the system.
Clearly, $H_{soc}$ does not conserve spin and,  although it does not induce a pure dephasing, it influences spin coherence through relaxation and energy exchange with the environment. 
For gate-defined neighbor quantum dots consists of AlGaAs/GaAs heterostructure, introduced in~\cite{QD14}, these parameters are given as $\beta{\simeq}4\alpha$.
In Fig.~\ref{fig:realistic} (d) the impact of SOC on the dynamics of the average fidelity in two operational modes of the transistor, in terms of coupling strength $\alpha$, is presented. Although both operational modes are suppressed, the obtainable average fidelities remain over $0.9$ for $\alpha{\leq}0.01t$.

\section{Experimental proposal}     
In typical quantum dot systems, on-site repulsion energy $U$ is on the order of $1$ meV, while inter-dot tunnel coupling $t$ and site-specific electrochemical potential $\varepsilon_{k}$ can be controlled precisely by metallic gate voltages~\cite{QD11,QD12,QD14}.
In a realistic quantum dot arrays with $L{=}3$ and $n{=}1$ electron, one can tune the parameters to have $t{=}0.02$ meV, $U{=}1$ meV ($U{/}t=50$), $V{=}0.1$ meV ($V{/}t=5$), $\varepsilon_{s}^{opt}{=}0.7$ meV ($\varepsilon_{s}^{opt}{/}t=35$), $\varepsilon_{open}^{opt}{=}0.3$ meV ($\varepsilon_{open}^{opt}{/}t=15$) for open and $\varepsilon_{closed}^{opt}{\geq}0.22$ meV ($\varepsilon_{closed}^{opt}{/}t\geq 11$) for closed transistors. 
These parameters result in $\overline{\mathcal{F}}_{open}(\tau^{opt}){>}0.98$ and $\overline{\mathcal{F}}_{closed}(2\tau^{opt}){>}0.99$ at optimal time $\tau^{opt}{\cong}14.68$ ns ($t\tau^{opt}{=}71$). The operation of the transistor is  much faster than decoherence time ${\sim}100 \mu s$ which allows for ${\sim}10{-}100$ subsequent operations.


\section{Conclusion}
We have proposed a low control QST for itinerant electrons to operate at the single electron level and control the flow of quantum information between distant qubits. The transistor can simply be controlled by external voltages which tune the charge configuration of electrons in the gate port. While the charge movement between the different ports of the transistor is suppressed by keeping them off resonant, the spin information can freely propagate between the source and the drain in the open mode. The performance shows higher fidelities and faster dynamics in comparison with the well-studied spin chain based state transfer scenarios. The switching between the open and the closed modes is fast which allows for several subsequent operation. The performance which is robust against several types of imperfections, makes this as potential practical application for quantum dot arrays.


\section{Acknowledgment} 
AB acknowledges support from the National Key R\&D Program of China (Grant No.2018YFA0306703) and National Science Foundation of China (grants No.12050410253 and No.92065115). SB thanks the
EPSRC grant for Non-ergodic quantum manipulation
EP/R029075/1.


\appendix* 
\setcounter{equation}{0}
\setcounter{figure}{0}
\setcounter{table}{0}
\renewcommand{\theequation}{A\arabic{equation}}
\renewcommand{\thefigure}{A\arabic{figure}}
\renewcommand{\thetable}{A\arabic{table}}

\section{}
In this section, we first describe our methodology for constructing the Hamiltonian matrices, then derive the ground states of $H_{gate}$ for both modes of operation, i.e., open and closed, analytically. In the next step, we analyze the mechanism of the transistor for rebuilding the initial state of the source on the drain qubit.    
For a transistor consists of the source, drain and a gate with $L{=}3$ sites, the basis specifying by the occupations of the electrons on the lattice and their spin projections is 
\begin{equation}\label{basis}
\vert\{n_{k\sigma}\}\rangle=\vert{n_{s\uparrow},n_{1\uparrow},n_{2\uparrow},n_{3\uparrow},n_{d\uparrow},n_{s\downarrow},n_{1\downarrow}
,n_{2\downarrow},n_{3\downarrow},n_{d\downarrow}}\rangle, 
\end{equation}
with $n_{k\sigma}{=}0$ or $1$ and subscript $s$ and $d$ for source and drain sites. 
For a gate with $n{=}1$ electron, $H_{gate}$ is a $9{\times}9$ matrix. Because of the conservation of the number and spin polarization, the gate Hamiltonian can be divided into two independent blocks with different spins $\uparrow$ and $\downarrow$ as
\begin{equation}\label{eq.h_gate}
H_{gate}=
\begin{bmatrix}
H_{gate}^{\uparrow} & 0 \\
0 & H_{gate}^{\downarrow} 
\end{bmatrix}, \quad \mathrm{with} \quad 
H_{gate}^{\sigma}=
\begin{bmatrix}
-\varepsilon_{1} & -t & 0 \\
-t & -\varepsilon_{2} & -t\\
0 & -t & -\varepsilon_{1} \\
\end{bmatrix}.
\end{equation}
Here the imposed mirror symmetry on the local potentials are  considered.
The eigensystem of each $H_{gate}^{\sigma}$ is obtained as  
\begin{eqnarray}\label{eq.eigen3}
\begin{aligned}
\lambda_{0,2}=\dfrac{1}{2}(-\varepsilon_{1}&-\varepsilon_{2} \mp \sqrt{8t^2+(\varepsilon_{1}-\varepsilon_{2})^2}),  \quad  
\lambda_{1}=-\varepsilon_{1},      
\\
\vert E_{0}\rangle &=(\vert O_{\sigma} \rangle - x_0\vert C_{\sigma} \rangle)/\sqrt{1+x_0^2},
\\ 
\vert E_{1}\rangle &= (-\vert {\sigma_{g}},0,0 \rangle + \vert 0,0,{\sigma_{g}}\rangle)/\sqrt{2}, 
\\ 
\vert E_{2}\rangle &=(\vert O_{\sigma} \rangle - x_2 \vert C_{\sigma} \rangle)/\sqrt{1+x_2^2},
\end{aligned}
\end{eqnarray}
where 
\begin{eqnarray}\label{eq:ground states}
\vert O_{\sigma}\rangle = (\vert{\sigma,0,0}\rangle{+}\vert{0,0,\sigma}\rangle)/\sqrt{2} \;\;\;\mathrm{and}\;\;\;
\vert C_{\sigma}\rangle = \vert{0,\sigma,0}\rangle
\end{eqnarray}
and $x_j{=}\dfrac{1}{t}(\varepsilon_{1}+\lambda_{j})$ with $j{=}0,2$.
Assuming $(\varepsilon_{1}-\varepsilon_{2})^2\gg 8t^2$, 
for $|\varepsilon_{1}|{>}|\varepsilon_{2}|$ 
straightforward calculations lead to 
$\lambda_{0}\simeq (-\varepsilon_{1}-\dfrac{2t^2}{\varepsilon_{1}-\varepsilon_{2}})$ 
and 
$\lambda_{2}\simeq (-\varepsilon_{2}+\dfrac{2t^2}{\varepsilon_{1}-\varepsilon_{2}})$ 
which in turn result in 
$\lambda_{0}\lesssim \lambda_{1}\ll \lambda_{2}$,
and  
$\vert E_{0}\rangle \simeq \vert O_{\sigma}\rangle$.
Therefore by setting the potential configuration as 
$|\varepsilon_{1}|{>}|\varepsilon_{2}|$, 
the ground state of $H_{gate}$ almost collapses on 
$\vert O_{\sigma}\rangle$.
Tuning the chemical potential as 
$|\varepsilon_{1}|{<}|\varepsilon_{2}|$ 
and analogues calculations provide 
$\lambda_{0}\simeq (-\varepsilon_{2}-\dfrac{2t^2}{\varepsilon_{2}-\varepsilon_{1}})$ 
and 
$\lambda_{2}\simeq (-\varepsilon_{1}+\dfrac{2t^2}{\varepsilon_{2}-\varepsilon_{1}})$.
Still 
$\lambda_{0}\lesssim \lambda_{1} \ll \lambda_{2}$, 
and the ground state of 
$H_{gate}^{\sigma}$ can be simplified to  
$\vert E_{0}\rangle \simeq \vert C_{\sigma}\rangle$.
These results show that how setting the potential landscape results in different $H_{gate}$ with different ground states that provides two desired open and closed operational modes of our transistor.


%

\end{document}